\documentclass[12pt,reqno]{amsart}

\usepackage{amssymb,amsmath}
\usepackage{graphicx,psfrag}

\theoremstyle{definition}
\newtheorem{example}{Example}

\title{Closed Timelike Curves in Relativistic Computation}
\author{Hajnal Andr\'eka,  Istv\'an N\'emeti and
  Gergely Sz\'ekely}
\begin{document}

\begin{abstract}
In this paper, 
we investigate the possibility of using closed
timelike curves (CTCs) in relativistic hypercomputation.  We 
introduce a wormhole based hypercomputation scenario which is free
from the common worries, such as the blueshift problem.  We also
discuss the physical reasonability of our scenario, and why we cannot
simply ignore the possibility of the existence of spacetimes
containing CTCs.
\end{abstract}
\maketitle

\section{Introduction}

G\"odel in 1949 
introduced a cosmological solution of Einstein's field equations in
which time travel is possible since it contains closed timelike curves
(CTCs), see \cite{Godel}.\footnote{G\" odel's spacetime is not the
  first in the literature that turned out to contain CTCs.}  Besides
G\"odel's rotating universe, there are several other interesting and
physically relevant spacetimes containing CTCs, such as Kerr-Newman
rotating black holes \cite[Prop.2.4.7]{ONeil} or Tipler's rotating
 cylinder \cite{Tipler} to mention only a few. For more
physically realistic spacetimes containing CTCs, see, e.g.,
\cite{Bonnor}.

There are several papers using CTCs to
design computers with higher computational power. For example, Brun
\cite{Brun} uses Novikov's principle of self-consistency, see, e.g.,
\cite[p.1916]{FMN}, to design algorithms solving the prime
factorization, NP-complete and even PSPACE-complete problems
efficiently.
There are other papers using Deutsch's causal consistency model
\cite{Deutsch} based on Everett's many world interpretation of
quantum mechanics to prove theorems about CTC based quantum
computation \cite{AW}, \cite{Bacon}.

These papers do not aim to challenge the physical
Church-Turing thesis by using the CTCs for hypercomputation (i.e.,
physical computational scenarios which are able to solve non-Turing
computable problems).

In the literature, the well-known concept of Malament-Hogarth
spacetime (MH-spacetime for short) is designed to capture those
spacetimes which are suitable for hypercomputation, see, e.g.,
\cite[\S 4.3]{earman}, \cite[\S 3]{Earman-Norton}, \cite{Hogarth},
\cite{HogarthPhD}. A spacetime is called a MH-spacetime if there is an
event (MH-event) whose causal past contains an infinitely long
timelike curve. The idea behind this definition is that a computer
traveling along this infinite path can compute forever and send a
signal at any time during its computation which reaches an observer
(programmer) before the MH-event.

Ignoring the potentially infinite space that the computer might
require to carry out an infinite computing task, we can argue that
hypercomputation can be implemented in any MH-spacetime as
follows. Let us take, for example, the non-Turing computable problem
of the decision of the consistency of Zermelo-Fraenkel set theory
(ZF). While the programmer moves along a finite worldline through the
MH-event, the computer checks all the proofs from the axioms of ZF one
by one looking for a contradiction. If the computer finds a
contradiction, let it send a signal to the programmer. It is easy to
see that the programmer does not get any signal if and only if no
contradiction can be derived from ZF. So if there is no signal before
the MH-event then the programmer learns that ZF is
consistent.\footnote{Let us note here that there is no contradiction
  with the fact that G\"odel's second incompleteness theorem implies
  that the consistency of ZF cannot be derived from the axioms of
  ZF. This is so because, if ZF is consistent, the above hypercomputational
  scenario does not prove the consistency by deriving it, but
  decides the question of consistency by a physical experiment.}

It is easy to show that every spacetime containing a CTC is a
MH-spacetime, see, e.g., \cite[Prop.1]{Manchak}.  So if the definition
of MH-spacetimes were our only criteria to accept a spacetime suitable
for hypercomputation, then here we could easily close our
investigation with the conclusion that hypercomputation can be
implemented in any spacetime containing CTCs.

Nevertheless, we do not stop here. In this paper, we introduce a
scenario that uses wormhole based CTCs for hypercomputation concerning
other aspects of spacetime needed for hypercomputation, such as the
potentially infinite space required by the computation.  We will show
that our construction has many advantages over the hypercomputational
scenarios of the literature, e.g., it is free from the common worries,
such as the blueshift problem.

There are several interesting papers dealing with the connection
of computation and the possibility of time travel. For example, Akl
\cite{Akl} investigates the issue of non-universality of CTC based
hypercomputation and Stannett \cite{Sta11} investigates the
possibility of P$\neq$NP in spacetimes containing special kinds of
CTCs.

For a survey on several physical and mathematical models related to
hypercomputation, see, e.g., Stannett \cite{Sta06}.

\section{Why don't we simply ban the existence of CTCs?}

Some physicists argue that time travel is too fancy to consider
spacetimes containing CTCs physically reasonable. They usually suggest
excluding them by assuming (as an axiom) that there are no CTCs in
physically reasonable spacetimes. However, this direct way of banning
 CTCs has a serious drawback. Namely, if we would like to
understand whether CTCs can or cannot occur in a ``physically
reasonable'' spacetime (e.g., in our universe), it is important not to
exclude these spacetimes by brute force (i.e., assuming that CTCs do
not exist).  As Monroe writes in his paper \cite{Monroe}: ``Thus,
causality assumptions, like Euclid's parallel postulate, risk closing
off interesting lines of investigation.''  In general, if we ban a
physical phenomenon by an axiom directly, we will not be able to investigate it
any more.  Therefore, we will never be able to (meaningfully) answer
the question why this phenomenon cannot occur. So if we assume that
CTCs cannot exist, the only thing we can say about why it is
impossible to travel back to the past is  because we have just
assumed it.\footnote{For more on why-type question in physics, see
  \cite{wtq}.}  In other words, if we simply ban CTCs by an axiom we
will never get any clue why they cannot occur in our universe or
whether they really cannot.

\label{p-ec}
It is interesting that finding natural axioms defining the physically
reasonable spacetimes is not an easy task at all. For example, the so
called energy conditions, which were a kind of natural way to exclude
the ``undesirable'' spacetimes, have come to seem less natural in
recent decades. Among other things, this is so because some of them simply
exclude the possibility of the accelerating expansion of our universe
(which has been discovered in 1998, see, e.g., \cite{riess}). So for
example, we cannot assume the ``natural'' strong energy condition if
we want to model our actual universe within general relativity. So the
strong energy condition is simply ``dead,'' but the weaker conditions
are also ``moribund'' for various reasons, see, e.g., \cite{BV}.  The
condition of being ``hole-free,'' introduced by Geroch, also turned
out to be too strong since Krasnikov has shown that even
Minkowski spacetime does not satisfy it \cite{K09}.

Spacetime theories consistent with CTCs are not only interesting
because they can be used to design hypercomputers, but because the
concept of time travel (independently of whether it is possible in our
world or not) is interesting by itself from the point of view of logic
since just like the Liar paradox it contains a kind of self-reference
(which is the basis of the grandfather paradox as well as the other
paradoxes of time travel).

\section{Problems with putting the computer to the CTC region}\label{sec-probl}

In using CTCs for hypercomputation, it is a natural idea sending the
computer back in time. However, if we put the computer exactly on the
CTC, the computer will go through the same chain of events. So we have
not created a hypercomputer. We have only created a time traveling
computer trapped in an infinite loop.

To avoid this kind of infinite loop of events, we can try to send the
computer back to a location close to its past self. However, if the
computer calculates long enough, there will be a (potentially) infinite
heap of computers which might turn into a black hole ruining the whole
project{ of hypercomputation}.  For example, if ZF is consistent,
a computer trying to decide its consistency will compute
forever. Hence, in this case, our heap of computers will be infinite.

To avoid creating a black hole, we have to send the computer back such
that the overall mass of the created computer heap does not increase
too fast with respect to the radius of the heap. We have to avoid that
the radius of the heap becomes smaller than the Schwarzschild radius
corresponding to the overall mass of the heap. The Schwarzschild
radius is proportional to the mass. So we can avoid creating a black
hole if we ensure that the mass of the heap of computers is
proportional to the radius of the heap.  If the size of the computer
does not increase, this task can be solved, e.g., by putting the spacelike
separated occurrences of the computer on a line with fixed distances
apart.

\begin{figure}
\psfrag{prog}[lt][lt]{Programmer}
\psfrag{comp}[rt][rt]{Computer}
\psfrag{MH}[lb][lb]{MH-event}
\psfrag{sign}[lb][lb]{signal}
\psfrag{t1}[tl][tl]{$t_1$}
\psfrag{t2}[tl][tl]{$t_2$}
\psfrag{Id}[l][l]{identify}
\psfrag{Time}[tl][tl]{Time}
\psfrag{bI}[r][r]{identify}
\includegraphics[keepaspectratio, width=0.9\textwidth]{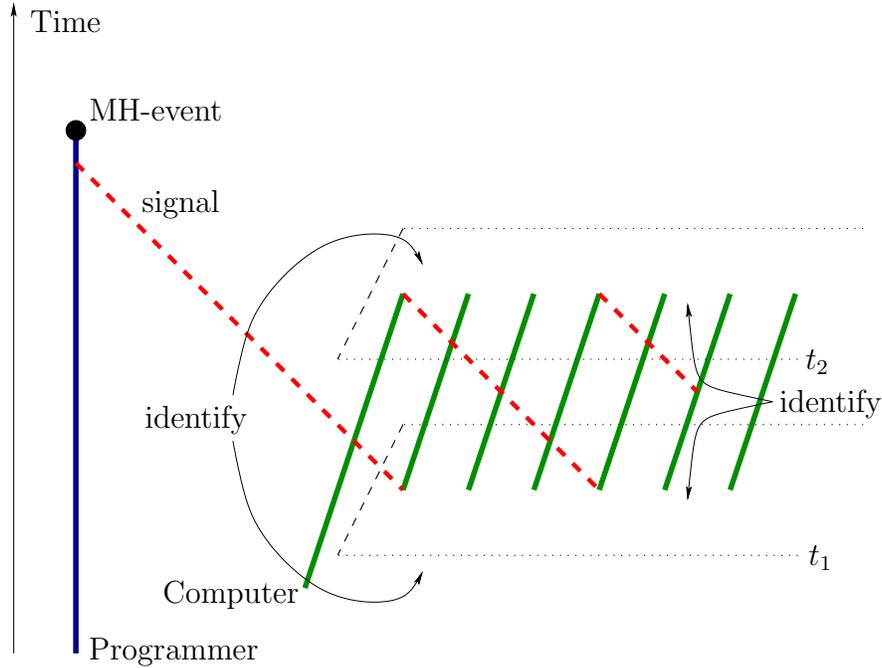}
\caption{\label{fig-c1} A 3 dimensional illustration of Example \ref{ex1}}
\end{figure}

Let us now create a four dimensional\footnote{All the examples
  of this paper can be constructed in any spacetime dimension grater
  than 2, and some of them, e.g., Example \ref{ex1}, can even be
  constructed in two dimension.} (toy) spacetime in which the
scenario above can be implemented.
\begin{example}\label{ex1}
Let us identify the lower and the upper parts of two horizontal half
hyperplanes, illustrated by two half spaces, in Minkowski spacetime,
see Figure~\ref{fig-c1}.\footnote{By this identification, we do
    not change the metric of the Minkowski spacetime just its
    topology. We make the identification in the following way. Let us
    first choose two horizontal half hyperplanes such that they
    intersect the same vertical lines. Let us first remove the edges
    of these half hyperplanes (these points will not be part of
    our new spacetime). Let $t_1<t_2$ be the time coordinates
    corresponding to the two half hyperplanes.  Let us connect the
    edges of the interval $[t_1, t_2)$; and similarly let us
      connect the edges of the half lines $(\infty,t_1)$ and
      $[t_2,\infty)$ in every vertical line intersecting the interior
        of the two half hyperplanes, i.e., every vertical line
        intersecting the interior of the half hyperplanes is replaced
        with a line and a circle. The circles merge into a 4
        dimensional half cylinder which will be the CTC region of the
        spacetime and the lines merge into a Minkowski half
        space. These two new regions is connected through the other
        Minkowski half space, which we have left unchanged.}
\end{example}

It is easy to see that our CTC based hypercomputation scenario can be
implemented in the spacetime of Example \ref{ex1} if the size of the
computer is fixed. However, fixing the size of the computer bounds its
computing power, too. So because of this bound, we have only created a
fast computer with limited computational power, but that is not a
hypercomputer.

It is highly probable that the MH-spacetime of Example \ref{ex1} does
not contain enough space for hypercomputation.  This is so because the
distance of the spacelike separated instances of the computers can
only be increased by increasing the velocity of the computer which is
bounded from above by the speed of light. 

\begin{figure}
\psfrag{prog}[l][l]{Programmer}
\psfrag{comp}[l][l]{Computer}
\psfrag{MH}{MH-event}
\psfrag{sign}{signal}
\includegraphics[keepaspectratio, width=0.8\textwidth]{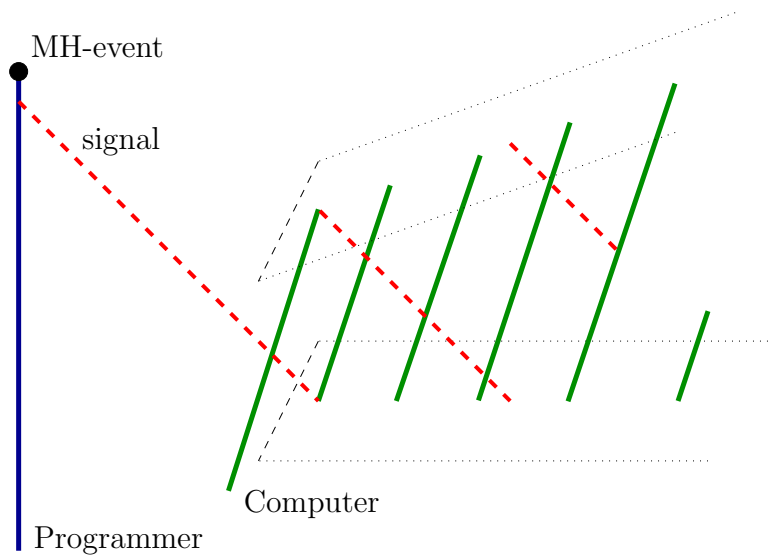}
\caption{\label{fig-c2} A 3 dimensional illustration of Example \ref{ex2}}
\end{figure}

We can overcome the distance problem of Example \ref{ex1}, by changing
it the following way.
\begin{example}\label{ex2} 
Let us identify, by the method used in Example \ref{ex1}, two
non parallel spacelike half hyperplanes such that the distance of the
spacelike separated instances of the computer increases without
changing its speed, see Figure~\ref{fig-c2}.
\end{example}

In Example \ref{ex2}, we have given a CTC based MH-spacetime in which
the (slow enough) increase of the mass of the computer does not risk
that the heap of computers will turn into a black hole.

If the size of the computer increases with time, we have to send it
back such that the distance of the spacelike separated instances of
the computer increases, too. By Examples \ref{ex1} and \ref{ex2}, we
have illustrated that whether this can or cannot be done depends on the
type of the CTC region we use.

It is clear that sending the computer back to the past over and over
again exaggerated the difficulty of increasing the size of the
computer without creating a black hole. This is so because we not only
have to be careful not to turn the computer itself into a black hole,
but we also have to ensure that the distances between its simultaneous
appearances increase fast enough. Without sending it back to the past,
we could easily ensure that the computer will never turn into a black
hole by increasing its size faster than its mass.

By the arguments above, it is likely that the computer does not have
enough space for hypercomputation in the spacetime of Example
\ref{ex1}. However, except if the spacetime is 2-dimensional, our
argument does not prove undoubtedly that the computer, by some
cunning trick, cannot use the infinite space orthogonal to its
movement without turning into a black hole.

So let us construct a (4-dimensional) MH-spacetime in which the
computer clearly does not have infinite space for hypercomputation.

\begin{example}\label{ex3} Let us replace the
half-spaces with 3-dimensional stripes having finite width and
identify them the same way as in Example \ref{ex1}.
\end{example}

\begin{figure}
\psfrag{prog}[l][l]{Programmer}
\psfrag{comp}[l][l]{Computer}
\psfrag{MH}{MH-event}
\psfrag{sign}{signal}
\psfrag{Fin}{finite distance}
\includegraphics[keepaspectratio, width=0.8\textwidth]{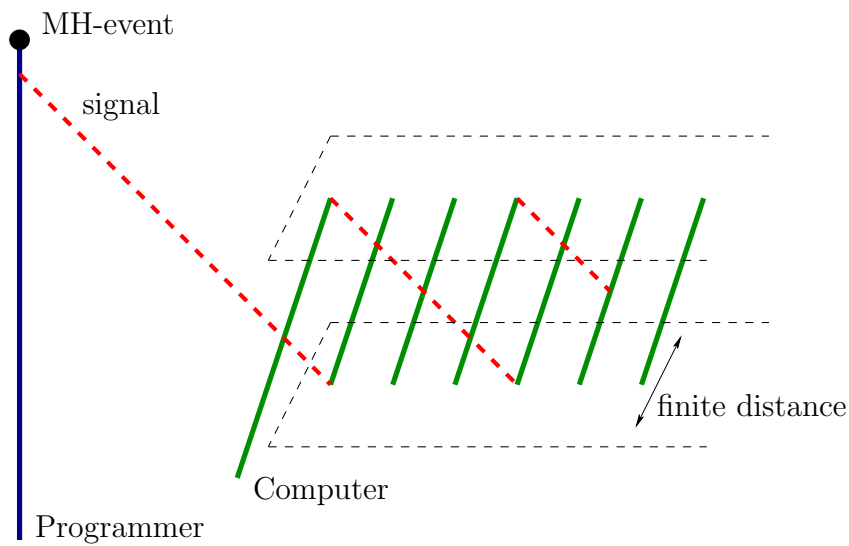}
\caption{\label{fig-c11} A 3 dimensional illustration of Example \ref{ex3} }
\end{figure}

Example \ref{ex3} is clearly a  MH-spacetime in
which hypercomputation is not possible due to the lack of space.
Our Example \ref{ex3}, shows the existence of MH-spacetimes, in which
hypercomputation cannot be implemented.  So (without regarding the
question of physical relevance) there are MH-spacetimes in which the
project of hypercomputation cannot be carried through. Therefore, the
concept of MH-spacetimes has to be refined if we want to capture the
concept of those (not necessarily physically reasonable) spacetimes in
which hypercomputation is possible.

Several papers, e.g., Etesi-N\'emeti \cite{Etesi-Nemeti} or
N\'emeti-D\'avid \cite{Nemeti-dgy} put great effort into showing that
in appropriate cosmological backgrounds (which are consistent with our
experimental data about our real universe) the Kerr-Newman rotating
black holes are physically relevant spacetimes suitable for
hypercomputation. For example, \cite{Nemeti-dgy} deals not only with
the issue of the potentially infinite space the computation requires
or the famous blueshift problem, but it also takes into account the
relevant quantum theoretical considerations, such as the evaporation
problem of black holes due to Hawking radiation.  However, these
papers do not try to introduce precise definitions to extend the
concept of MH-spacetimes to capture the spacetimes which are suitable
for relativistic hypercomputation.

Manchak \cite{Manchak} introduces some explicitly defined properties
(such as, signal reliability condition or finite acceleration
condition) of MH-spacetimes to make them more realistic for
hypercomputation. However, he does not list any property ensuring the
potentially infinite space required by the computation in the
desirable properties. Moreover, his construction does not contain
infinite space for the computer to compute. 

Obviously, relativistic hypercomputation is an infinitely
expensive project. So at first it may sound strange trying to lower its
cost. However, even if it is infinitely expensive, it is not the same
if the maintenance of the computer costs \$1 per century or
\$1,000,000 per minute. Let us note here that from the point of view
of the hypercomputation project the computer can be arbitrarily slow
since it has infinite time to compute. So if the computer travels
along a geodesic, its yearly maintenance cost can be quite cheap if it
is calculates slow enough.

According to a conjecture of Andr\'eka-N\'emeti-W{\" u}trich, in every
spacetime where the CTCs are created by some kind of rotation, the
CTCs have to counter rotate with the rotation creating them. This
conjecture is valid in all the well-known spacetimes where CTCs are
created by rotation \cite{ANW}.\footnote{For a visual explanation of
  this counter rotation effect in the case of G\"odel's universe, see
  \cite[\S 6]{NMAA}.} To counter rotate with a rotating mass the
computer has to be accelerated.\footnote{For example, Malament showed
  that the total integrated acceleration of any CTC in G\" odel
  spacetime is at least $\ln(2+\sqrt{5})$ \cite{malament}.}  As the
mass (size of its data storage) of the computer increases during its
calculation it becomes more and more expensive (it takes more and more
energy) to accelerate it. And that can make it difficult (if not
impossible) to keep the yearly cost (required energy) of the computer
bounded above.

\section{A CTC based relativistic hypercomputer that actually works without problems}

Here we are going to present a relativistic hypercomputational
scenario based on CTCs that works without the problems presented in
Section~\ref{sec-probl}.  All the problems above were generated by
sending the computer back in time.  So let's try to design a
hypercomputer using CTCs without sending the computer back to the
past.

Is it possible to exploit the CTCs without using it for sending the
computer back to the past?  Yes, it is possible. The key idea is to
send only the final result of the computation back to the past.

Let us first create, by using the ``cut and paste'' method of our
previous examples, a spacetime containing a CTC in which
hypercomputation is possible without sending the computer back to the
past.

\begin{figure}
\psfrag{prog}[l][l]{Programmer}
\psfrag{comp}[l][l]{Computer}
\psfrag{A}{A}
\psfrag{B}{B}
\psfrag{MH}{MH-event}
\psfrag{sign}{signal}
\includegraphics[keepaspectratio, width=0.8\textwidth]{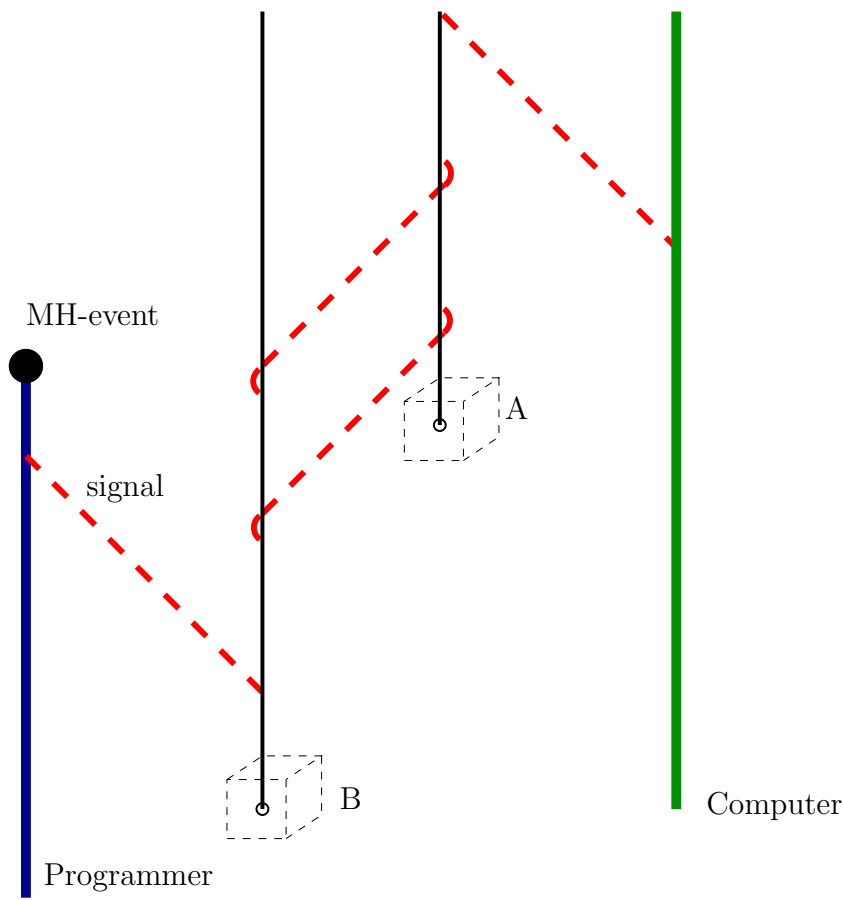}
\caption{\label{fig-c3}  Illustration of Example \ref{ex4} }
\end{figure}

\begin{example}\label{ex4}
Let us identify, by the method used in Example \ref{ex1}, the sides of two ``stationary'' boxes in the Minkowski
spacetime such that identification of one of them begins (much)
earlier than the other, see Figure~\ref{fig-c3}.
\end{example}

By Example \ref{ex4}, we have created a kind of wormhole such that one of its
mouths is in the past relative to its other mouth.

Let us now see how a hypercomputational scenario (e.g., decision
of consistency of ZF) can be implemented in this spacetime.  It is
clear that, in this spacetime, the computer has enough space and time
to compute remaining outside the CTC region.
 
Let the computer send a signal to mouth A of the wormhole if it
derives a contradiction from ZF.  By the identification, the signal
comes out from mouth B earlier. Now let some device send (reflect) the
signal back to mouth A. If the time delay between mouths A and B is
more than the time that the signal has to take to reach mouth A from
mouth B, the signal enters earlier to mouth A. Repeating this cycle
the signal comes out from mouth B in the far past (only a little later
than the initialization of the computer) where the programmer waits for the
information. So if the computer derives a contradiction from ZF the
programmer receives the signal. And if ZF is consistent (i.e., no
contradiction can be derived from it), the programmer does not receive
any signal and learns that ZF is consistent.

It is an important feature of the construction that the computer can
send a signal back to the past at any time during its infinite
computation.  It is clear that not every spacetime containing a CTC
can be used to implement this scenario.  For example, the spacetimes
of Examples \ref{ex1}, \ref{ex2} and \ref{ex3} are not suitable for a
computer avoiding the CTC region to send signals back to the past.

It is clear that in our scenario, based on Example \ref{ex4}, sending
only the final signal back to the past using the CTC region eliminates
all the problems of Section~\ref{sec-probl}. Moreover, the famous
blueshift problem (see, e.g., \cite{Etesi-Nemeti}, \cite{Nemeti-dgy})
of hypercomputation based on Kerr-Newman black holes simply does
not show up in this scenario.

\section{Physically more realistic  spacetimes using CTCs for hypercomputation} 

In Example \ref{ex4}, we have used the ``cut and paste'' method for creating
our spacetime, which is a good method for creating counter examples to
show that some logical implications about spacetimes do not
hold. However, this method does not have any other physical relevance.

\begin{figure}
\psfrag{prog}[l][l]{Programmer}
\psfrag{comp}[l][l]{Computer}
\psfrag{A}{A}
\psfrag{B}{B}
\psfrag{MH}{MH-event}
\psfrag{Wormhole}{Wormhole}
\psfrag{sign}{signal}
\includegraphics[keepaspectratio, width=0.8\textwidth]{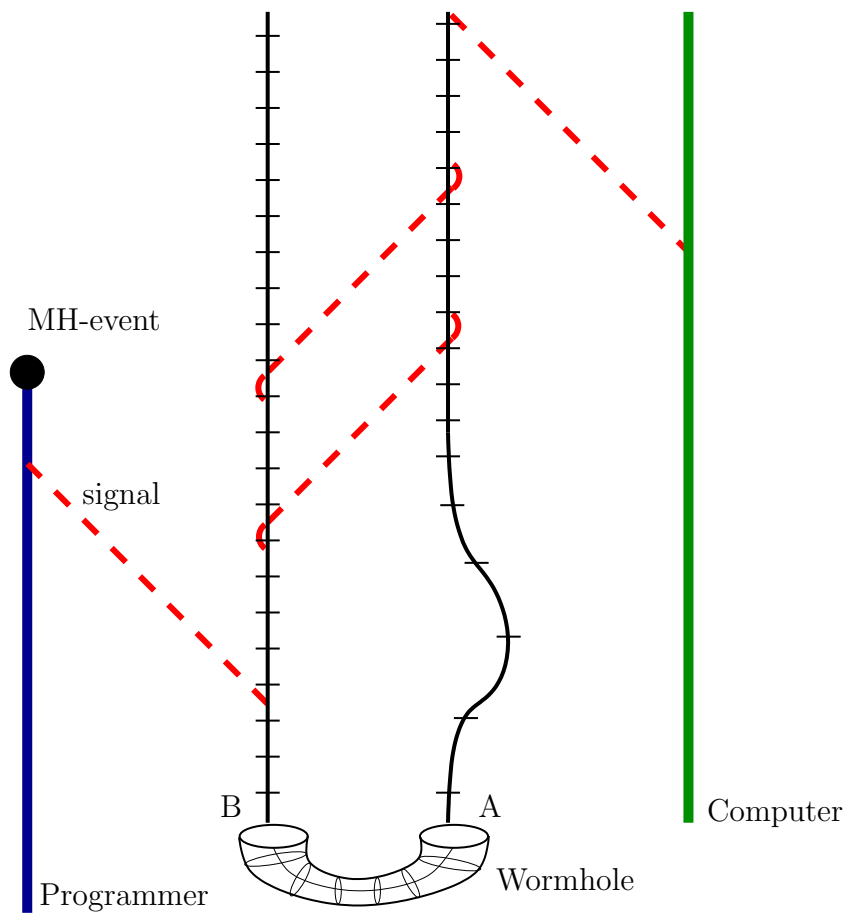}
\caption{\label{fig-c5}  Illustration of Example \ref{ex5}} 
\end{figure}

By using the twin paradox theorem of relativity theory, we can
slow-down time at one of the mouths of the wormhole, see, e.g.,
\cite[\S B]{FMN}, \cite[\S I]{MT}, \cite{MTY}.  This slow-down creates a spacetime
similar to our spacetime of Example \ref{ex4} in which the above
hypercomputational scenario can be realized the same way. So we can
create a CTC region similar to that of our ``cut and paste'' toy model above by
using the twin paradox and wormholes.
\begin{example}\label{ex5}
We take a spacelike wormhole and accelerate one\footnote{The
      mouth of the wormhole can be accelerated, e.g., by accelerating some ordinary
      matter before the mouth which will dragging it by gravitation, see \cite[p.146]{Novikov}.}
  of its mouths in such a way that after the acceleration the time dilation,
  caused by the twin paradox effect, between the two mouths is grater than the
  amount of time a signal needs to travel from one mouth to the other,
  see Fig.~\ref{fig-c5}. 
\end{example}

In the construction of our wormhole based spacetime, we have only used
the twin paradox theorem of relativity theory and wormholes, whose
existence is consistent with the theory of general
relativity. 
Moreover, wormholes have become more realistic after the
  discovery of the acceleration of the expansion of our universe, see, e.g., \cite{BV}.


\section{On the physical reasonability of this setup}

The rotating black hole based hypercomputation sends the programmer
into a black hole and leaves the computer outside to compute forever,
see, e.g., \cite{Etesi-Nemeti}, \cite{Nemeti-dgy}. A nice feature of
our wormhole based hypercomputation as opposed to the rotating
black-hole scenarios is that the programmer can remain at home and
does not have to travel into a rotating black hole. A drawback is that
we do not know whether there are wormholes in our universe or how to
create them.  While there is strong observational evidence of the
existence of huge rotating black holes (see, e.g., \cite{MGH}) which
are ideal for hypercomputation, we do not have such strong
experimental evidence for the existence of wormholes. However, there
are attempts to detect wormholes, and some astronomical objects
seem promising, see, e.g., \cite{KNS}. Of course, if there are no
(non-quantum size) wormholes in our universe, there is still hope that
one day we will be able to create some, e.g., by enlarging some
quantum wormholes \cite[\S H]{MT}, \cite{MTY} or by using the Casimir
effect \cite{V-gia}.  However, these are only speculations right now.

So wormholes are only speculative objects for the time being.  To put
some optimism to the end of this section, let us close it by the
following claim of Visser written in 1997: ``The good news about
Lorentzian wormholes is that after about ten years of hard work we
cannot prove that they don't exist.''\cite{V-gia}.\footnote{That
    is, so far there is no observation of a wormhole (or laboratory
    experiment providing one), but at least it is difficult to
    disprove their existence based on our spacetime theories.}

\section{Repeatability of the hypercomputational experiment}

So we might decide whether ZF set theory is consistent or not via our
hypercomputer. If the answer is yes, probably we would like to use
another hypercomputer to check whether a stronger axiom system (e.g.,
ZF together with some large cardinality assumption) is still
consistent. If we found out that ZF is inconsistent, obviously we
would like to use a hypercomputer to check the consistency of some
weaker set theory (e.g., the theory of hereditary finite sets). Not to
mention that there are other interesting non-Turing computable
questions besides the decision of the consistency of set theories.  So
it would be nice if the hypercomputational experiment would be
repeatable.

The hypercomputational scenarios are typically not repeatable or their
repeatability strongly depends on things that we do not know or cannot
influence, such as what the programmer finds inside the rotating
black hole he has jumped into.

What about the repeatability of our wormhole based
hypercomputation?  If there are wormholes, then probably there are
more than one or if we can create wormholes, then probably we can
create more than only one of them. So, if we can have one, it is
plausible that we can have several from the key part of our
hypercomputer. What about the infinite spacetime that the
hypercomputer consumes during its computation?  It is reasonable to
assume that our universe is potentially infinite (since its expansion
accelerates right now and we have absolutely no reason to think that
this fact will change in the future).  If we have infinite space for
the first computer, then we will have enough space for the second, the
third, etc. computers if we use it wisely. The trick is simple: use
only 1/2,000,000 of the infinite space for the first computer (it is
enough since it is still infinite) use 1/4,000,000 for the second
computer, 1/8,000,000 for the third, etc. Then we will never pollute
more than 1/1,000,000 of our universe with the electronic waste of our
hypercomputers.  So it is reasonable that the wormhole based
hypercomputation is repeatable.

\section{faster than light motion and closed timelike curves}

It is a common belief that faster than light motion, which is possible
in (1+1) dimension even for observers, see, e.g., \cite[\S
  2.7]{pezsgo}, \cite[\S 2.7]{Madarasz}, entails CTCs.  In the case of
(1+3) dimension, the fact that no observer can move faster than light
can be derived (in the sense of mathematical logic) from a streamlined
axiom system of special relativity called \textsf{SpceRel}, see
\cite[Theorem 2.1]{Synthese}. In the proof of Theorem 2.1 of
  \cite{Synthese}, it was strongly used that the dimension of space is
  grater than that of time. This fact and that there can be observers
  moving faster than light in the case of (1+1) dimension motivates us to 
conjecture that there is a natural generalization of \textsf{SpecRel}
for the case of (3+3) dimension in which faster than light motion is
allowed for observers.

If an
observer sends out a faster than light signal, this signal
travels backwards in time according to some observers moving relative to
him. This fact suggests that, if observers can move faster than light,
then they can travel back in time. So the above common belief about
CTCs and faster than light motion seems reasonable.

However, it is not true. Using the axiomatic method it is possible
to show that faster than light motion of observers does not imply the
possibility of time travel by itself.  Moreover, time travel is
possible by using faster than light observers only if it is possible
without them.\footnote{This result is based on a joint research of Mike
Stannett (University of Sheffield) and our research group led by
Hajnal Andr\'eka and Istv\'an N\'emeti (R\'enyi
Institute) \cite{AMNSSz}.}

Why are we interested in (1+1)-dimensional and (3+3)-dimensional
spacetime theories when we are apparently living in a (1+3)-dimensional
one?  Because we are logicians and therefore we are interested in
logical connections of the possible axioms (basic assumptions) of our
theories including the ones concerning the space and time
dimensions. Besides this, we also would like to understand all the
possible universes (that could have been created) and not just our
sole universe we live in.

\section{Concluding remarks}
We have seen that sending a computer back in time is a kind of awkward
and problematic way to use CTCs to create hypercomputers. The idea of
sending only the final result of the computer back in time gives us a
much more convenient way to utilize the CTCs.  We have shown using
wormholes a spacetime containing CTCs that can be utilized for
hypercomputation by sending only signals back. We have seen that
except that the existence of wormholes is less well-supported by
experimental evidence than that of rotating black holes, wormhole
based hypercomputation has several advantages over the black hole
based version.

It would be interesting to create and logically analyze (in the spirit
of \cite{logst}, \cite{Synthese}, \cite{Madarasz}, \cite{Szphd}) an
axiomatic theory of relativistic computation containing not only basic
concepts required by the spacetime theory but also the concepts needed
to formulate the scenario of hypercomputation in the language of the
theory. The same way as the axiomatic investigation leads to a deeper
(more logical) understanding of relativity theories, it would lead to
a better understanding of the theory of relativistic
hypercomputation.

We do not claim that the story ends here or that all the questions of
wormhole based hypercomputation are answered here. On the contrary, we
think that there remain several interesting unanswered questions and
our main motivation is to arouse the interest about the subject of
wormhole based hypercomputation as a possible rival of the black hole
based one.

\section{Acknowledgment}
We are grateful to Mike Stannett for his valuable suggestions and
comments. We are also grateful to Pavel Pudl\'ak for  inspiring
discussions on the subject.  This research is supported by the
  Hungarian Scientific Research Fund for basic research grants
  No.~T81188 and No.~PD84093.

\bigskip\bigskip
\noindent {\sc H.~Andr\'eka,
I. N\'emeti, G. Sz{\'e}kely\\} Alfr{\'e}d R{\'e}nyi
Institute of Mathematics \\ of the Hungarian Academy of Sciences\\
Budapest P.O.
Box 127, H-1364 Hungary\\ {\tt andreka@renyi.hu,
nemeti@renyi.hu,
  turms@renyi.hu}

\end{document}